\documentstyle[12pt,aaspp]{article}
\tighten
\eqsecnum
\slugcomment{Accepted to the Astronomical Journal}
\begin{document}

\title{Shock-Excited Maser Emission from Supernova Remnants: 
G32.8$-$0.1, G337.8$-$0.1, G346.6$-$0.2, and the HB3/W3 Complex}
\author{Barron Koralesky \altaffilmark{1}}
 \affil{Department of Astronomy, University of Minnesota, 116 Church 
Street SE, Minneapolis, MN 55455}
\author{D. A. Frail\altaffilmark{2}, W. M. Goss\altaffilmark{3}, M. 
J. Claussen\altaffilmark{4}}
 \affil{NRAO/AOC, P.O. Box O, Socorro, NM 87081}
\author{and A. J. Green\altaffilmark{5}}
 \affil{University of Sydney, School of Physics, Sydney, NSW 2006, 
Australia}

	\altaffiltext{1}{e-mail: barron@astro.spa.umn.edu}
	\altaffiltext{2}{e-mail: dfrail@nrao.edu}
	\altaffiltext{3}{e-mail: mgoss@nrao.edu}
	\altaffiltext{4}{e-mail: mclausse@nrao.edu}
	\altaffiltext{5}{e-mail: agreen@physics.usyd.edu.au}

\begin{abstract}

We present the results of VLA observations in the ground-state 
hydroxyl (OH) transition at 1720 MHz toward 20 supernova remnants 
(SNRs).  We detect compact emission from four objects.  For three of 
these objects (G32.8$-$0.1, G337.8$-$0.1, and G346.6$-$0.2), we argue 
that the emission results from masers which are shock-excited due to 
the interaction of the SNR and an adjacent molecular cloud.  We 
observe a characteristic Zeeman profile in the Stokes V spectrum, 
which allows us to derive a magnetic field of 1.5 and 1.7 mG for 
G32.8$-$0.1 and G346.6$-$0.2, respectively.  The velocity of the 
masers also allows us to determine a kinematic distance to the SNR. 
Our criteria for a maser to be associated with an SNR along the line 
of sight are that the position and velocity of the maser and SNR must 
agree, and the OH(1720) emission must be unaccompanied by other OH 
lines.

\end{abstract}


\section{Introduction}

Supernova remnants (SNRs) are a primary source of energy and heavy 
elements in the ISM. While this is well accepted, it is difficult to 
demonstrate specific instances of interaction between the SNR and cold 
dense gas in the ISM. Massive stars evolve quickly and are thought to 
influence the molecular clouds (MCs) from which they formed (see 
Elmegreen 1998 for a review).  Circumstantial evidence, such as 
morphological signatures, has been the primary evidence of this 
interaction (e.g. Routledge et al.  1991, Landecker et al.  1987).  
Interpretation of such observations is inconclusive because of 
confusion due to projections along the line of sight.  Thus there is a 
need for an indicator of SNR-MC interaction.

In a series of recent papers, we have utilized the 1720 MHz line of 
the hydroxyl molecule (OH) as a diagnostic for such an interaction.  
The ground state of the OH molecule has four transitions: the main 
lines at 1665.4018 and 1667.3590 MHz and the satellite lines at 
1612.231 and 1720.530 MHz.  Main line OH masers are commonly found 
near compact H\thinspace{II} regions, while 1612 MHz OH masers are 
found in the circumstellar shells of evolved stars.  Goss \& Robinson 
(1968) first detected the satellite 1720 MHz OH maser emission toward 
the SNRs W28 and W44.  They found strong narrow-line 1720 MHz emission 
and broad-line 1612, 1665, and 1667 MHz {\it absorption}, which 
distinguished the 1720 MHz masers from those found in other 
environments.

The remnants W28 and W44 later became prototypes for SNRs interacting 
with molecular clouds.  Frail, Goss, \& Slysh (1994) reobserved W28, 
discovering 26 maser spots, and argued that the 1720 MHz radiation is 
maser emission due to a strong inversion of the line from collisional 
excitation by H$_{2}$ behind the SNR shock.  The location of the maser 
spots and the physical characteristics of the SNR and MC are 
consistent with the shock excitation model of Elitzur (1976).

In addition to being an unambiguous sign of SNR-MC interaction, OH 
masers can tell us more about the host SNR and its physical 
characteristics.  In systems where the velocity has been measured by 
other means, there is strong agreement between the velocity of the 
masers and the systemic velocity of the remnant.  This allows a 
kinematic distance to be derived.  Polarimetric observations reveal 
Zeeman profiles in the Stokes V signal which allow us to measure the 
magnetic field in several of the maser spots (Claussen et al.  1997).  
We can already loosely constrain the temperature and density of the 
environment of the maser via Elitzur's (1976) theory.  Furthermore, in 
the future it may be possible to use maser theory to infer accurate 
physical parameters of the postshock gas from the OH masers.

The initial success with W28 prompted Frail et al.  (1996) to perform 
a survey for 1720 MHz OH maser emission toward 66 SNRs, using the 
Green Bank and Parkes telescopes.  A total of 19 detections were made.  
Follow-up observations with an interferometer were made to distinguish 
between compact maser emission ($\theta < 1''$) and spatially extended 
thermal line emission.  This resulted in the detection of OH(1720 MHz) 
masers toward five new SNRs.  This survey work has continued toward 
the Galactic Center (e.g. Yusef-Zadeh et al.  1996) and in the 
Galactic plane (Green et al.  1997).

Green et al.  (1997) reported that of the $\sim$160 SNRs searched for 
OH emission, 17 have yielded maser emission, or 10\% of the sample.  
However, they report higher detection rates for remnants within the 
``molecular ring'' of the Galaxy.  Many of these individual SNRs have 
several maser spots.  For instance, W28, W44 and IC 443 have 41, 25, 
and 6 masers, respectively (Claussen et al.  1997).  The 10\% 
detection rate in a random sample of Galactic SNRs suggests that more 
SNR-MC interactions have yet to be discovered in the remaining 
systems.

In the Green et al.  survey of 75 SNRs, there were 33 detections of 
OH. However, 16 were not re-observed with an interferometer to 
determine whether the detections result from compact maser emission or 
extended Galactic thermal emission.  The goal of the present study was 
to determine which of the remaining detections visible from the 
VLA\footnote{The National Radio Astronomy Observatory is a facility of 
the National Science Foundation operated under cooperative agreement 
by Associated Universities, Inc.} were due to shock-excited OH maser 
emission.  In addition, we also observed several other sources which 
had not been sampled by the Green et al.  survey, but were likely 
cases for SNR-MC interaction.  Two of these were Cas A and CTB109.  
Also, we have noticed a strong correlation between OH maser detections 
and members of J. Rho's (1996) thermal composite SNRs.  This may 
suggest molecular cloud interactions as an origin for this class of 
SNRs.  Therefore, we observed three SNRs from this category (3C396, 
CTB104A, and HB3).
 
\section{Observations}

The observations were made at the VLA on 25 October and 19 December 
1997 in the DnC and D configurations, respectively.  A bandwidth of 
780 kHz was used, divided into 127 spectral channels using the on-line 
Hanning smoothing.  This yielded both L and R hands of circular 
polarization with a velocity resolution of 1.1 km s$^{-1}$ (6.1 kHz 
per channel) and a velocity coverage of 120 km s$^{-1}$.  The band was 
centered at 1720.53 MHz, offset by the single-dish line emission 
velocity or the systemic velocity of the remnant (when known).  
Table~\ref{D-obs-table} lists the pointing centers for all 
observations along with the central velocities for the observations.

Follow-up observations were made on 5, 6, and 13 February 1998 during 
the move time between D and A configurations.  For the first two days, 
we had 24 antennas in the D configuration and we observed 
G\thinspace{32.8$-$0.1} and G\thinspace{346.6$-$0.2}.  For the third 
day, we used 13 antennas which were in the A configuration and HB3 was 
observed.  A bandwidth of 195 kHz was used, divided into 127 spectral 
channels using on-line Hanning smoothing.  This yielded parallel 
handed circular polarizations with a velocity resolution of 0.53 km 
s$^{-1}$ (1.5 kHz per channel) and a velocity coverage of 34 km 
s$^{-1}$.  During these observations, we also observed in the main 
line transitions at 1665 and 1667 MHz with the same spectral 
resolution.  We centered, both spatially and spectrally, on the masers 
detected during our 1997 observations.

After the VLA observations, we became aware of an Australia Telescope 
Compact Array (ATCA) observation of one of the southern SNRs detected 
in OH(1720 MHz) by Frail et al.  (1996).  In a survey of star-forming 
regions in both satellite lines of OH, Caswell et al.  (1998) 
serendipitously observed the SNR G\thinspace{337.8$-$0.1} in one 
field.  This data was acquired on 1997 November 27 in the 6C 
configuration with a 4 MHz bandwidth and 2048 channels, yielding 0.68 
km s$^{-1}$ velocity resolution in the orthogonal terms of linear 
polarization.

All image processing was done using NRAO's AIPS package.  The 
continuum data was fitted via a least-squares method and subtracted.  
Three-dimensional image cubes were generated for all pointings in 
Stokes I, then CLEANed using the AIPS task IMAGR. This resulted in 
average rms noise levels of 5-10 mJy beam$^{-1}$.  Stokes V cubes were 
made for pointings where masers were detected.

\begin{figure}
	\vbox to7.0in{\rule{0pt}{2.6in}}
	\includegraphics{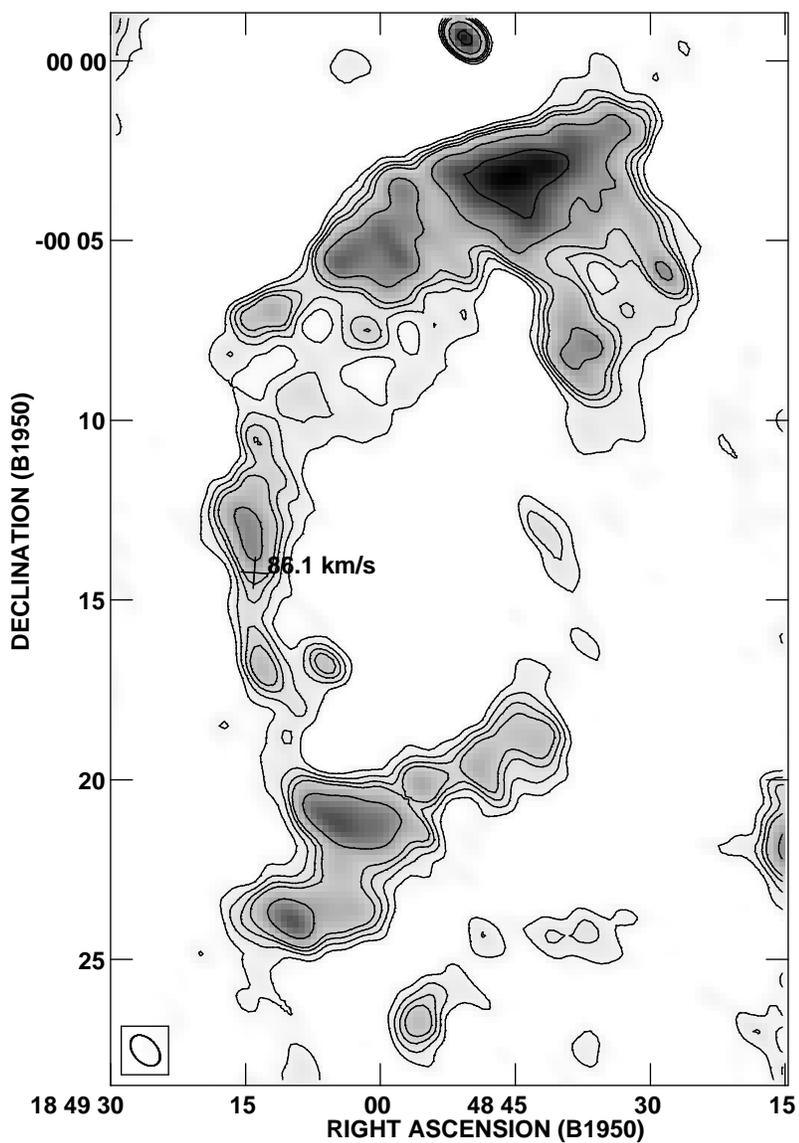}
\caption{Radio continuum image of G32.8$-$0.1 at 1.7 GHz from the current
data.  The position of the maser is marked as a cross on the image and the
velocity is noted.  The size and position angle of the cross are those
of the Gaussian fit.  Continuum contour levels are 3, 6, 12, 24, and 48 mJy
beam$^{-1}$ and the beamsize is 59\arcsec $\times$ 39\arcsec, as shown
in the lower left.  {\it For published version of this figure, see
http://ast1.spa.umn.edu/barron/research.html}}
\label{G32.8-fig}
\end{figure}

\begin{figure}
	\vbox to7.0in{\rule{0pt}{2.6in}}
	\includegraphics{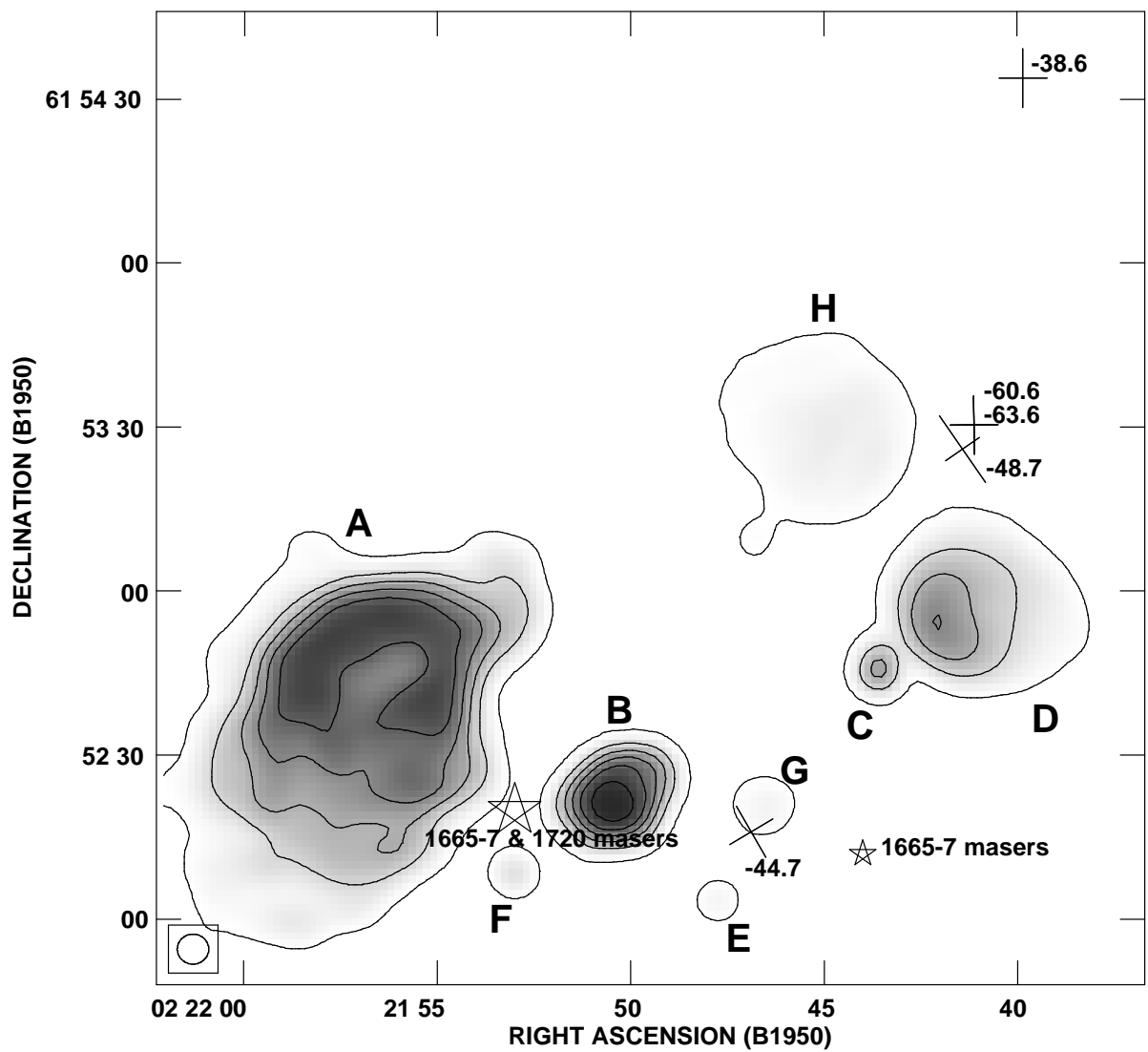}
\caption{Radio continuum image of W3 at 1.4 GHz from Kantharia, 
Anantharamaiah, \& Goss (1998), with resolution of $\sim$5.5\arcsec \ and
individual components labeled.  HB\thinspace{3} would be located off the 
map to the northwest. The positions of the masers are marked as crosses on 
the image and the velocity is noted.  The size and position angle of
the cross are those of the Gaussian fit.  Positions of groupings of 
1665 and 1667 MHz masers  are marked with a star and labeled as 1665-7.  
Contour levels are 15, 90, 165, 240, 315, and 390 mJy beam$^{-1}$ and
the continuum beamsize is 5.74\arcsec  $\times$ 5.48\arcsec, as shown
in the lower left.} 
\label{W3-fig}
\end{figure}

\begin{figure}
	\vbox to7.0in{\rule{0pt}{2.6in}}
	\includegraphics{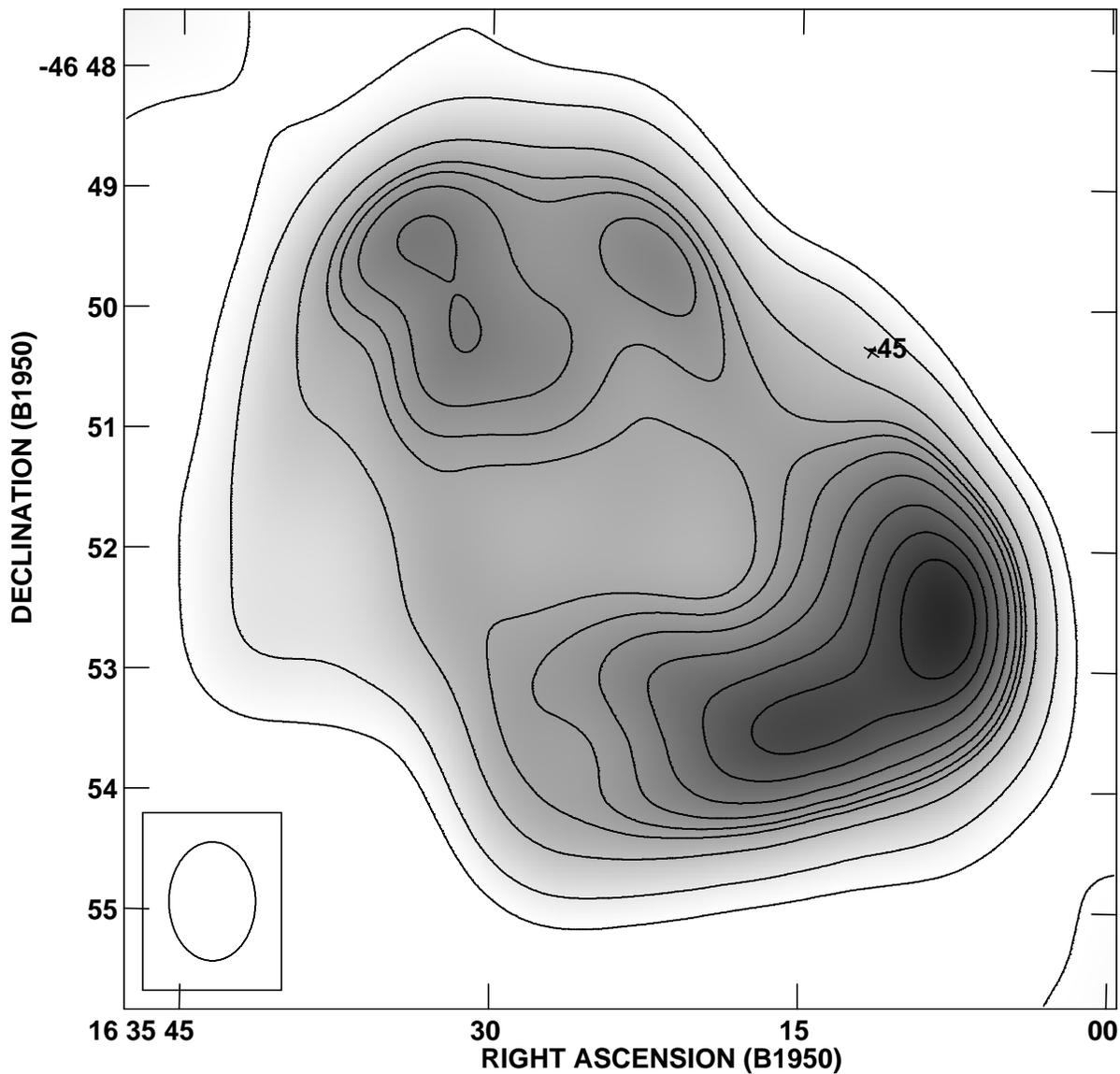}
\caption{Radio continuum image of G337.8$-$0.1 at 843 MHz from 
Whiteoak \& Green (1996).  The position of the maser is marked as a cross on
the image and the velocity is noted.  The size and position angle of
the cross are those of the Gaussian fit.  Contour levels are 55, 140,
290, 380, 420, 470, 540, 630, 710, and 780 mJy beam$^{-1}$ and the
continuum beamsize is 59\arcsec $\times$ 43\arcsec, as shown in the
lower left. {\it For published version of this figure, see
http://ast1.spa.umn.edu/barron/research.html}} 
\label{G337.8-fig}
\end{figure}

\begin{figure}
	\vbox to7.0in{\rule{0pt}{2.6in}}
	\includegraphics{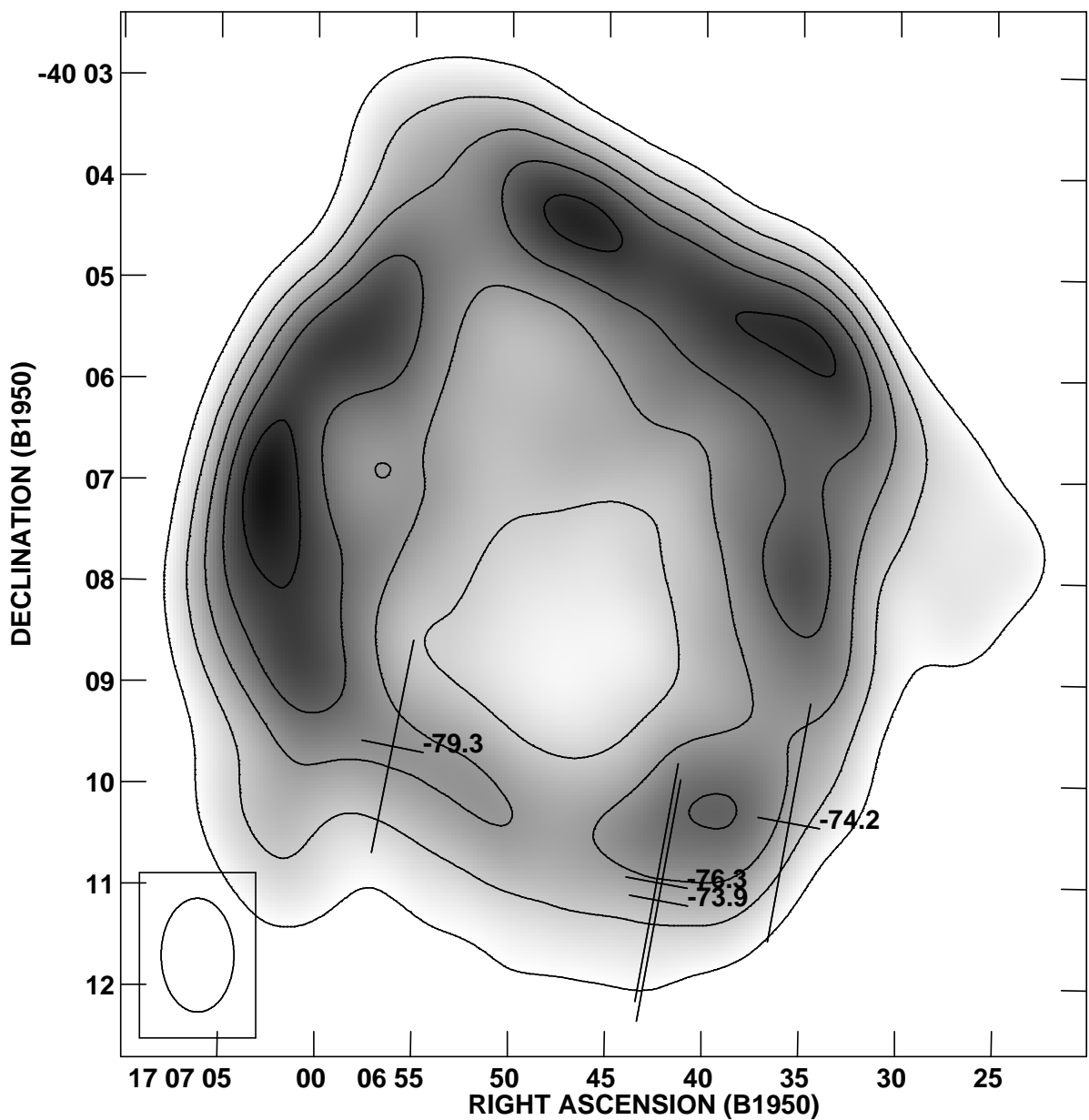}
\caption{Radio continuum image of G346.6$-$0.2 at 843 MHz.  The 
positions of the four masers are marked as crosses on the image and 
the velocities are noted.  The size and position angle of the crosses 
are those of the Gaussian fits.  Contour levels are 50, 100, 150, 
200, and 250 mJy beam$^{-1}$ and the continuum beamsize is 67\arcsec $\times$ 
43\arcsec, as shown in the lower left. }{\it For published version of this figure, see
http://ast1.spa.umn.edu/barron/research.html}
\label{G346.6-fig}
\end{figure}

Positional errors are roughly $0.1-0.2$\arcsec, or perhaps worse in 
accordance with the usual formula for a positional error $\Delta\theta 
= \pm(\theta/2)(S/N)^{-1}$.  The sizes of the beams are also 
represented by the size of the crosses in Figures~\ref{G32.8-fig}, 
\ref{W3-fig}, \ref{G337.8-fig}, and \ref{G346.6-fig}.

\section{Results}

With the completion of the present work, all of the northern SNRs 
($\delta>-45^\circ$) detected by Green et al.  (1997) with single-dish 
detections of the 1720 MHz line have been followed up with the VLA. 
Several southern SNRs await confirmation.  We observed 17 SNRs, 
detecting compact maser emission toward two of them.  This 
confirmation rate of 10\% is similar to previous surveys (Frail et al.  
1996, Green et al.  1997, Yusef-Zadeh et al.  1996).  There are now a 
total of 19 SNRs towards which we have detected OH(1720 MHz) masers.  
Below we discuss each of our recent detections, reviewing what is 
known about the individual SNR and whether our maser detections imply 
a physical association between the maser and the remnant.

\subsection{G\thinspace{32.8$-$0.1} (Kes\thinspace{78})}

Kes\thinspace{78} is an elongated shell-type SNR approximately 
20\arcmin\ (N-S) by 10\arcmin\ (E-W).  Although the quality of our 1.7 
GHz continuum image in Figure~\ref{G32.8-fig} has limited sensitivity, 
the remnant does resemble that of earlier lower resolution images (see 
Kassim 1992 and references therein).  We detected a single maser spot 
at V$_{\rm{LSR}}$=+86.1 km s$^{-1}$, along the eastern edge of the 
SNR. Caswell et al.  (1983) identified an H$_2$O maser near the center 
of Kes\thinspace{78} coincident with main-line OH maser emission in 
this direction.  This maser likely originates in an unrelated compact 
HII region along the line of sight, since it is 7\arcmin\ away (18$^h$ 
46$^m$ 10\fs5, $-$00\arcdeg 15\arcmin 36\arcsec (B1950)) from our 
OH(1720 MHz) detection and has V$_{\rm{LSR}}$=+33 km s$^{-1}$.

\begin{figure}
	\vbox to7.0in{\rule{0pt}{2.6in}}
	\includegraphics{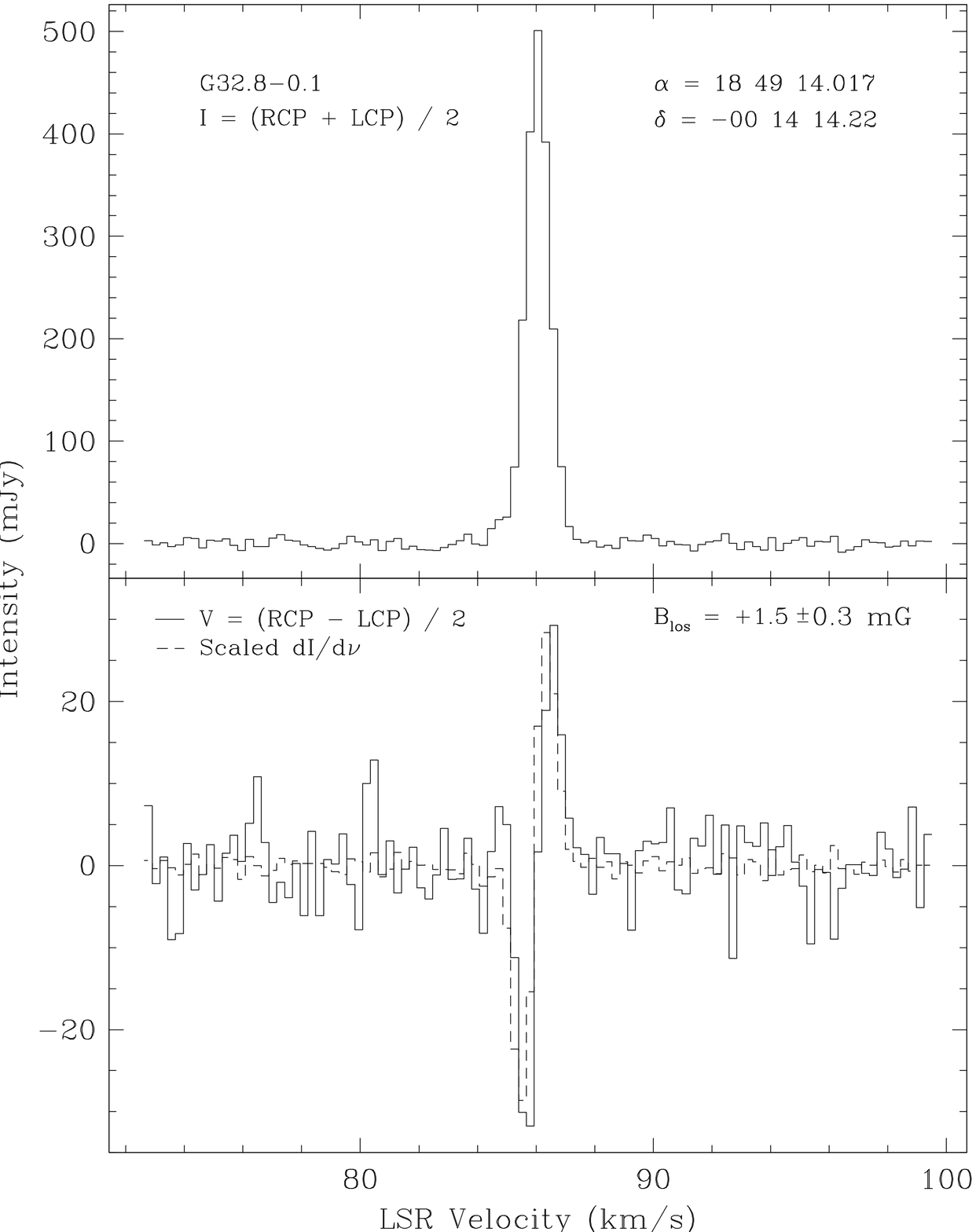}
\caption{Stokes I and V spectra for the maser feature in G32.8$-$0.1.  
The derivative of I with respect to frequency (dashed line) scaled by 
the magnetic field strength is shown plotted with the V spectrum.}
\label{MC-G32}
\end{figure}

We also detected a significant signal in Stokes V, with the classical 
antisymmetric S-shaped profile about the peak in the Stokes I. Such a 
profile is characteristic of Zeeman splitting in the presence of a 
magnetic field (Heiles et al.  1993).  As we have done for earlier 
efforts (Yusef-Zadeh et al.  1996, Claussen et al.  1997), we fit the 
derivative of the Stokes I to the Stokes V profile for this maser 
source and derived a line-of-sight measurement of the magnetic field.  
The result of this fit is shown in Figure~\ref{MC-G32}.  Good 
agreement can be obtained with the V profile for B$_{los}=+1.5 \pm 
0.3$ mG. This value lies nearly centered between those which we 
measured earlier for the SNRs W\thinspace{44} and W\thinspace{28} 
(B$_{los}\simeq$0.2 mG) and Sgr A East (B$_{los}\simeq$3 mG).

If we make the assumption that the shock is transverse to the 
line-of-sight, then the velocity of the masers represents the LSR 
velocity of the SNR (Frail et al.  1996, Claussen et al.  1997), we 
can use the rotation curve of the Galaxy (Fich, Blitz \& Stark 1989) 
to derive the kinematic distance to the remnant.  Without any other 
distance discriminator, +86.1 km s$^{-1}$ corresponds to either a near 
distance of 5.5 kpc or a far distance of 8.8 kpc.  The tangent point 
in this direction (assuming a distance to the Galactic center of 8.5 
kpc) is at 7.1 kpc with V$_{\rm{LSR}}\simeq{+100}$ km s$^{-1}$.

\subsection{The HB3/W3 Complex}

The SNR HB3 is a large, evolved remnant embedded in a star-forming 
region which contains the W3 H\thinspace{II} complex.  Routledge et 
al.  (1991) found W3 and HB3 to lie at $-43$ km s$^{-1}$ and estimate 
HB3 to be 60 $\times$ 80 pc in diameter.  Routledge et al.  made a 
case that HB3 and W3 lie at the same distance due to their discovery 
of a bright region of CO emission associated with W3 that is partially 
surrounded by continuum emission from HB3, all occurring at the 
systemic velocity.  While it appears that HB3 is interacting with 
molecular gas in which W3 is embedded, there is no evidence for a 
direct interaction of W3 and HB3.  W3 lies to the SE of HB3, and their 
centers are separated by about 1\arcdeg, or about 40 pc at the 
distance of the system (Landecker et al.  1987; Normandeau, Taylor, \& 
Dewdney 1997).

We found six 1720 MHz masers (see Table~\ref{W3-maser-table}) without 
corresponding 1665/7 main line emission in the direction of HB3 and 
W3.  We also reduced archival VLA data, taken on January 20 and 23, 
1989, when the array was in the A configuration.  The pointing center 
was toward the W3 Main complex of HII regions, and all four 
ground-state OH transitions were observed.  We were able to map the 
1720 MHz OH masers toward W3(OH), the main part of W3, and toward the 
southeast part of HB3.  The velocity coverage was from -80 to -10 km/s 
with respect to the local standard of rest.  These earlier 
observations confirmed four out of the six 1720 MHz masers reported in 
Table~\ref{W3-maser-table}.  Also, none of the four had associated 
main line emission to the level of about 50 mJy/beam (3 $\sigma$).

All but one of these can be immediately ruled out as being due to the 
HB3-W3 interaction because of position (see Figure~\ref{W3-fig}).  The 
remaining northernmost maser is close to the interaction region in the 
southeast quadrant of HB3.  Spatially, it is near the peak in the 
$^{12}$CO and, at $-38.6$ km s$^{-1}$, it coincides with the velocity 
of the bulk of the molecular material between $-36$ to $-45.9$ km 
s$^{-1}$.  However, comparing with Routledge et al., we found the 
maser lies outside of the 408 MHz continuum emission of HB3.  This led 
us to conclude that none of the masers seen in this direction are 
collisionally excited by the shock of HB3, and, instead, they are most 
likely associated with the compact H\thinspace{II} regions of W3.

\subsection{G\thinspace{337.8$-$0.1} (Kes\thinspace{41})}

A single maser was detected toward Kes\thinspace{41}, shown 
superimposed on the 843 MHz continuum image from Whiteoak \& Green 
(1996) in Figure~\ref{G337.8-fig}.  It lies on the northwestern edge 
between the twin ``caps'' of enhanced radio emission that define the 
distorted 6\arcmin\ by 4\arcmin\ shell of this SNR. Caswell et al.  
(1975) have measured HI absorption toward this SNR and see evidence 
for absorption by gas up to the tangent point at 7.9 kpc (for 
R$_{\odot}$=8.5 kpc) at V$_{\rm{LSR}}\simeq{-130}$ km s$^{-1}$.  If we 
accept this (the quality of the HI spectrum is not high), then it 
implies that Kes\thinspace{41} lies beyond the tangent point.  In this 
case, the velocity of the maser ($-$45 km s$^{-1}$) gives a kinematic 
distance of 12.3 kpc, making Kes\thinspace{41} a modest-sized remnant 
with a diameter of 20$\times$15 pc.

\subsection{G\thinspace{346.6$-$0.2}}

G\thinspace{346.6$-$0.2} is a little-studied remnant.  The image we 
show in Figure~\ref{G346.6-fig} is from the 843 MHz MOST survey of 
Whiteoak \& Green (1996) and, although it is well characterized as a 
circular shell of 8\arcmin\ diameter, there is pronounced flattening 
along the northwestern side.  This prompted Dubner et al.  (1993) to 
suggest that the SNR is interacting with the surrounding medium.  A 
total of five masers were detected toward this SNR. They are all 
located along the southern edge of the SNR and scatter within a few km 
s$^{-1}$ about a mean V$_{\rm{LSR}}\simeq{-76.0}$ km s$^{-1}$.  At 
this velocity, the kinematic distance to the masers (and hence the 
SNR) yields a near value of 5.5 kpc and a far value of 11 kpc.  The 
tangent point in this direction is at 8.3 kpc with 
V$_{\rm{LSR}}\simeq{-170}$ km s$^{-1}$.

\begin{figure}
	\vbox to7.0in{\rule{0pt}{2.6in}}
	\includegraphics{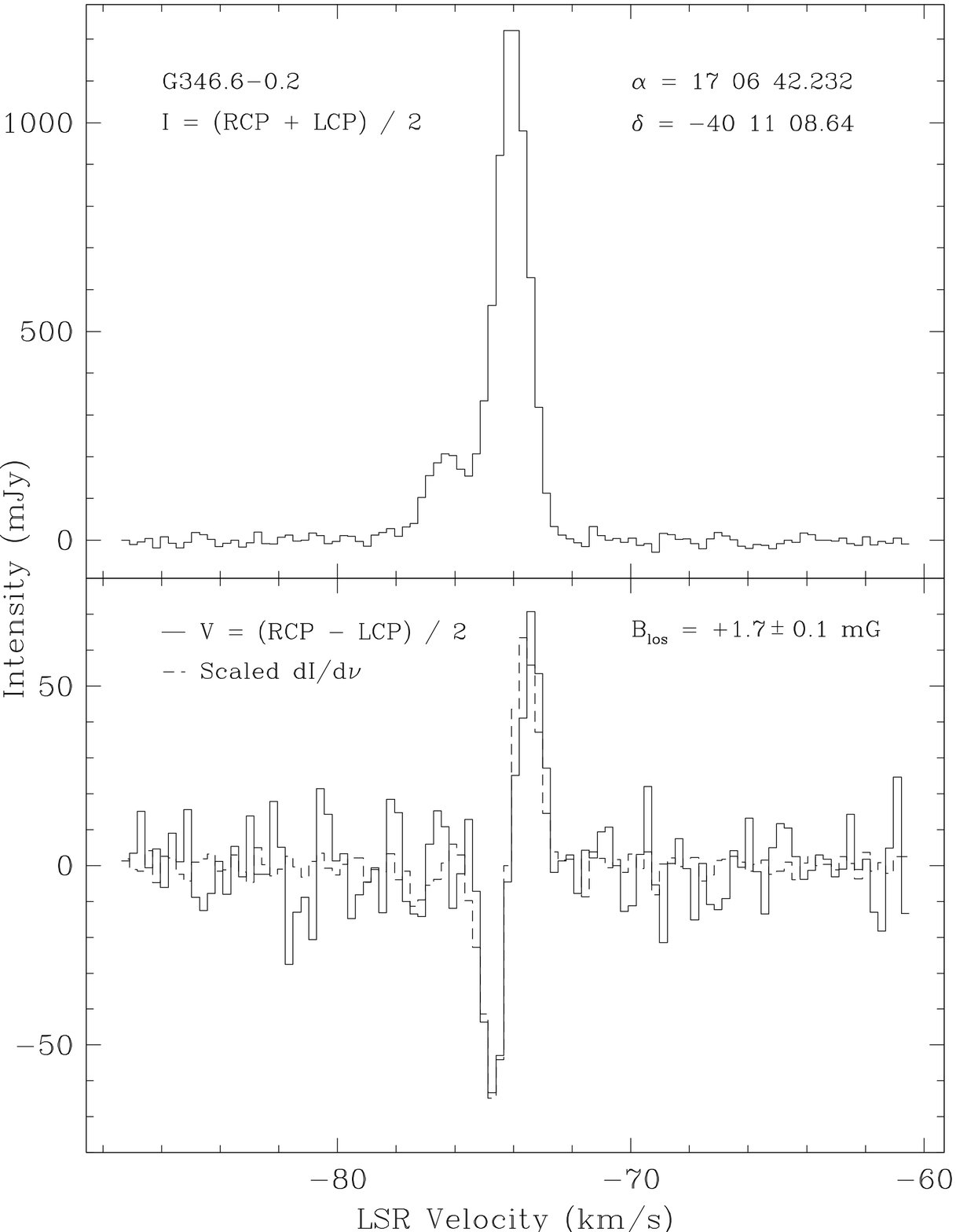}
\caption{Stokes I and V spectra for the brightest maser feature in 
G32.8$-$0.2.  The derivative of I with respect to frequency (dashed 
line) scaled by the magnetic field strength is shown plotted with the 
V spectrum.}
\label{MC-G346}
\end{figure}

In Figure~\ref{MC-G346} we show the Stokes I and V for the brightest 
maser line toward G\thinspace{346.6$-$0.2}.  The second component that 
can be seen in this figure is the weak line at $-$76.2 km s$^{-1}$.  
Significant V flux is only detected toward the strong maser.  
Following the same procedure as used for Kes\thinspace{78}, we derived 
B$_{los}=+1.7 \pm 0.1$ mG.

\section{Discussion}

As in previous surveys of this kind (Frail et al.  1996, Green et al.  
1997), we will argue below that the OH(1720 MHz) masers that we have 
detected in G\thinspace{346.6$-$0.2} and G\thinspace{32.8$-$0.1} are 
directly associated with the SNRs themselves.  From continued work of 
this kind it has become clear that these SNR OH(1720 MHz) masers 
represent a distinct class of OH masers in our Galaxy, separate from 
OH masers in star-forming regions and evolved stars.  Moreover, 
evidence is steadily accumulating that the SNR OH(1720 MHz) masers are 
collisionally excited by H$_2$ molecules and are therefore a good 
tracer of SNR interactions with molecular clouds.

Observationally there are several important signatures that we look 
for in establishing whether a particular maser detection is associated 
with a SNR along the line of sight.  First, we require close agreement 
in both position and velocity for the OH(1720 MHz) masers and the SNR. 
The masers should be found interior to, or along the radio continuum 
edges of, the SNR. For the three associations we have advocated here 
(Kes\thinspace{78}, Kes\thinspace{41} and G\thinspace{346.6$-$0.2}) 
this is certainly the case.  Other potential associations like HB3 and 
G\thinspace{27.4+0.0} (Green et al.  1997) are rejected on this basis.  
Velocity information can also be extremely useful but it is often 
difficult to obtain.  For example, in this work an HI absorption 
spectrum was available only toward Kes\thinspace{41}, yielding just a 
lower limit to the LSR velocity of the SNR. It has been noted in 
several instances where the LSR velocity of the SNR is known, that the 
maser velocities lie within a few km s$^{-1}$ of this value (Claussen 
et al.  1997, Green et al.  1997, Frail et al.  1996).  Such close 
agreement is taken to imply that the masers originate in shocks 
transverse to the line of sight, and we have used this fact to derive 
kinematic distances to SNRs with no prior velocity information.

A second important observational constraint is that we require that 
any OH(1720 MHz) maser detection be unaccompanied by main line 
emission or a 1612 MHz detection.  More specifically, Goss \& Robinson 
(1968) noted that for W\thinspace{28} and W\thinspace{44} both the 
main lines and 1612 MHz were seen in {\it absorption} against the 
radio continuum from the SNRs when the OH(1720 MHz) was seen in {\it 
emission}.  While the long baselines of the VLA used in our 
confirmation observations are necessary to identify the compact maser 
emission, they preclude OH absorption measurements.  Therefore, we 
often search in emission for the main lines of OH toward SNRs where 
there is any question of an association.  For example, the main lines 
proved very useful in understanding the OH(1720 MHz) detections toward 
HB3 and W\thinspace{51C}, both of which are located in complicated 
star forming complexes.  Furthermore, we did not find main line 
emission in G\thinspace{346.6$-$0.2} and G\thinspace{32.8$-$0.1} at 
the locations of our OH(1720) masers.  This criterion, although not 
infallible, is a useful discriminant.  Empirically it is known that 
OH(1720 MHz) emission is nearly non-existent toward stellar OH masers 
(there is the singular case of V\thinspace{1057}\thinspace{Cygni}, 
which is highly variable in the 1720 MHz line (see Lo and Bechis 1973 
\& 1974, Andersson et al.  1979, Winnberg et al.  1981)), and that 
OH(1720 MHz) masers in compact HII regions are almost always 
accompanied by main line and/or 1612 MHz emission (Gaume \& Mutel 
1987).

For those OH(1720 MHz) masers identified with SNRs using the criteria 
outlined above, there is ample evidence that these SNRs are undergoing 
an interaction with an adjacent molecular cloud and that the OH(1720 
MHz) masers are collisionally excited by H$_2$ molecules heated by the 
passage of the shock.  For the sample as a whole, Green et al.  (1997) 
noted that the SNRs with OH(1720 MHz) masers trace the distribution 
(in position and velocity) of molecular clouds in our Galaxy.  For 
individual remnants there exist morphological, kinematic and chemical 
signatures for molecular shocks.  Examples include the arcs and shells 
of molecular gas seen toward 3C\thinspace{391} and 
G\thinspace{359.1$-$0.5} (Yusef-Zadeh et al.  1996, Wilner et al.  
1998), the molecular cooling line of [OI] at 63 $\mu$m which peaks 
toward the masers in W\thinspace{44} and 3C\thinspace{391} (Reach \& 
Rho 1996), and broad lines ($\Delta V\sim{10-50}$ km s$^{-1}$) of CO 
and other mm/submm lines at the location of masers toward 
W\thinspace{28}, W\thinspace{44}, W\thinspace{51C}, and 
3C\thinspace{391} (Wootten 1977, 1981, Koo \& Moon 1997, Frail \& 
Mitchell 1998).  Although no published searches have been made for any 
sign of molecular shocks from Kes\thinspace{78}, Kes\thinspace{41}, or 
G\thinspace{346.6$-$0.2}, the inference can be drawn that SNRs with 
OH(1720 MHz) masers are prime candidates for such studies.

It should be stressed that while the {\it detection} of a OH(1720 MHz) 
maser spot in an SNR is strong evidence of a shock interaction with a 
molecular cloud, the converse is not true.  According to Elitzur 
(1976), a strong inversion of the OH molecule at 1720 MHz works under 
a limited range of range of kinetic temperatures (25$\leq$T$_k\leq200$ 
K) and H$_2$ gas densities (10$^3$ cm$^{-3}\leq{n} \leq 10^5$ 
cm$^{-3}$).  Moreover, the maser process requires large column 
densities of OH with low velocity dispersion in order to produce 
coherent amplification.  Thus, the {\it absence} of OH(1720 MHz) 
masers toward an SNR is not proof against an interaction, only that 
the conditions in the remnant are not favorable for their formation.  
For example, in this work we obtained null detections of OH(1720 MHz) 
toward Cas A and G\thinspace{109.1$-$1.0}, both of which have some 
evidence of an interaction with an adjacent molecular cloud (Tatematsu 
et al.  1990; Keohane, Rudnick, \& Anderson 1996; Koralesky \& Rudnick 
1998).  Likewise, it is worth noting that OH(1720 MHz) masers are not 
the sole province of SNRs.  Although 1720 MHz masers are less common 
in HII regions than their main line counterparts (Gaume and Mutel 
1987) they do exist, and in some cases they have no accompanying main 
line or 1612 MHz emission.  The isolated 1720 MHz detections that we 
made in W3 (\S{3}) are one such example.  Another is the bipolar 
outflow HII region S\thinspace{106} with the main lines detected in 
absorption and bright OH(1720 MHz) maser emission (Loushin 1989).  
More recent modeling of the excitation of the OH molecule by Pavlakis 
\& Kylafis (1996a, b) demonstrates that strong inversions at 1720 MHz 
are possible with radiative pumping at the densities and strong 
infrared radiation associated with compact HII regions.  Elitzur \& de 
Jong (1978) further showed that OH masers can arise in the interface 
zone between the shock and ionization front driven by the HII region.  
This underscores the need to apply the selection criteria outlined 
above in order to determine whether a maser detection belongs to the 
class of HII regions or SNRs.

We wish to thank J. Caswell for alerting us to 
G\thinspace{337.8$-$0.1} and for making his data available.  B. K.'s 
work was supported, in part, by NASA GSRP NGT-5-50123 and NSF grant 
AST-96-19438 to the University of Minnesota.  The Molonglo Observatory 
Synthesis Telescope is operated by the University of Sydney with 
support from the Science Foundation within the School of Physics and 
the Australian Research Council.  M. G. would like to thank R. D. 
Ekers and J. B. Whiteoak for their support during a sabbatical visit 
to the Australia Telescope National Facility in 1995.  B. K. would 
like to thank L. Rudnick for discussion and advice during this study.

\bigskip
{\bf References}
\bigskip

Andersson, C., Johansson, L. E. B., Winnberg, A., \& Goss, W. M. 1979, A\&A, 80, 260

Caswell, J. L. et al. 1998, in prep. 

Caswell, J. L., Murray, J. D., Roger, R. S., Cole, D. J., \& Cooke, D. J. 1975, A\&A, 45, 239 

Caswell, J. L., Batchelor, R. A., Forster, J. R., Wellington, K. J. 
1983, Aust. J. Phys. 36, 443   

Claussen, M.J., Frail, D.A., Goss, W.M., Gaume, R.A. 1997, ApJ, 489, 143

Dubner, G., Moffett, D. A., Goss, W. M., \& Winkler, P. F. 1993 AJ, 105, 2251 

Elitzur, M. 1976, ApJ, 203, 124 

Elitzur, M., \& de Jong, T. 1978, A\&A, 67, 323 

Elmegreen, B. G. 1998, in Origins of Galaxies, Stars, Planets, and 
Life, edited by C. E. Woodward, H.A. Thronson, \& M. Shull, Astronomical 
Society of the Pacific Conference Series, in press

Fich, M., Blitz, L., \& Stark, A. A. 1989, ApJ, 342, 272

Frail, D.A., Goss, W.M., \& Slysh, V.I. 1994, ApJ, 424, L111 

Frail, D. A., Goss, W. M., Reynoso, E. M., Giacani, E. B., Goss, W. M., \& Dubner, G. 1996, AJ, 111, 1651 

Frail, D. A., \& Mitchell 1998, ApJ, submitted

Gaume, R.A., \& Mutel, R.L. 1987, ApJS, 65, 193

Goss, W.M., \& Robinson, B.J. 1968, Astrophys. Lett., 2, 81

Green, A.J., Frail, D.A., Goss, W.M., \& Otrupcek, R. 1997,  AJ, 
114, 2058

Heiles, C., Goodman, A. A., McKee, C. F., \& Zweibel, E. G. 1993, in Protostars and Planets, ed E. H. Levy \&  J. 
I. Lunine (Tucson: Univ. Arizona Press), 279

Kantharia, N. G., Anantharamaiah, K. R. \& Goss, W. M. 1998,
ApJ, in press

Kassim, N. E. 1992 AJ, 103, 943 

Keohane, J.W., Rudnick, L., \& Anderson, M.C. 1996, ApJ, 466, 
309

Koo, B.-C., \& Moon, D.-S. 1997, ApJ, 475, 194

Koralesky, B., \& Rudnick, L. 1998, in prep.

Landecker, T. L., Dewdney, P. E., Vaneldik, J. F., \& Routledge, D. 1987, AJ, 94, 111

Lo, K. Y., \& Bechis, K. P. 1973, ApJ, 185, L71

Lo, K. Y., \& Bechis, K. P. 1974, ApJ, 190, L125

Loushin, R. S. 1989, PhD Thesis, Univ. Ill.

Normandeau, M., Taylor, A. R., \& Dewdney, P. E. 1997, ApJS, 108, 279

Pavlakis, K. G., \& Kylafis, N. D. 1996a, ApJ 467, 300

Pavlakis, K. G., \& Kylafis, N. D. 1996b, ApJ 467, 309

Reach, W. T., \& Rho, J. 1996, 315, 277

Rho, J. 1996, PhD thesis, Univ. of MD

Routledge, D. 1991, A\&A, 247, 529

Tatematsu, K., Fukui, Y., Iwata, T., Seward, F. D., \& Nakano, M. 1990, ApJ, 351, 157

Wilner, D. J., Reynolds, S. P., \& Moffett, D. A. 1998, AJ, 115, 247

Whiteoak, J. B. Z., \& Green, A. J. 1996, A\&AS, 118, 329

Winnberg, A., Graham, D., Walmsley, C. M., \& Booth, R. S. 1981, A\&A, 93, 79

Wootten, A. 1977, ApJ, 216, 440

Wootten, A. 1981, ApJ, 245, 105

Yusef-Zadeh, F., Roberts, D. A., Goss, W. M., Frail, D. A., \& Green, A. J. 1996, \apj, 466, L25

\begin{table}
\small
\begin{center}

\begin{tabular}{|c|c|c|c|c|c|}
\multicolumn{6}{c}{\sc Table~\ref{D-obs-table} -- Summary of December 1997 D and DnC Configuration 1720 MHz Observations}
\\
\hline
\hline
Name        & RA(1950)    & Dec(1950) & $V_{LSR}$ & Common Name & Notes\tablenotemark{\dagger} \\
     & (h m s) & ($\arcdeg$ $\arcmin$ $\arcsec$) & km s$^{-1}$ & & \\
\hline
G13.3-1.3A  & 18 14 39.0   & $-$18 09 30.0 &  $+30$ & & D \\
G13.3-1.3B  & 18 13 38.5   & $-$18 35 48.0 &  $+30$ & & D \\
G32.8-0.1   & 18 48 50.0   & $-$00 12 00.0 &  $+90$ & Kes78 & D \\
G33.6+0.1   & 18 50 15.0   & $+$00 37 00.0 &  $+90$ & Kes79 & D \\
G39.2-0.3   & 19 01 40.0   & $+$05 23 00.0 &  $+70$ & 3C396 & R \\
G93.7-0.2A  & 21 29 00.0   & $+$51 05 00.0 &  $-45$ & CTB104A & R \\
G93.7-0.2B  & 21 25 00.0   & $+$50 55 00.0 &  $-45$ &  & R \\
G109.1-1.0  & 22 59 00.0   & $+$58 45 00.0 &  $-50$ & CTB109 & \\
G111.7-2.1  & 23 21 10.0   & $+$58 32 00.0 &  $-40$ & CasA & \\
G132.7+1.3A & 02 16 00.0   & $+$61 45 00.0 &  $-40$ & HB3 & R \\
G132.7+1.3B & 02 19 00.0   & $+$61 50 00.0 &  $-40$ &  & R \\
G189.1+3.0  & 06 14 04.0   & $+$22 22 40.0 &   $ 0$ & IC443 & D \\
G340.6+0.3  & 16 44 05.0   & $-$44 29 00.0 &  $-100$& & D \\
G342.0-0.2  & 16 51 15.0   & $-$43 48 00.0 &  $ 0$  & & D \\
G341.9-0.3  & 16 51 25.0   & $-$43 56 00.0 &  $ 0$  & & D \\
G343.1-0.7A & 16 55 23.65  & $-$42 54 35.0 &  $-70$ & & D \\
G343.1-0.7B & 16 56 45.9   & $-$43 24 35.0 &  $-70$ & & D \\
G346.6-0.2  & 17 06 50.0   & $-$40 07 00.0 &  $-70$ & & D \\
G348.5-0.0  & 17 12 00.0   & $-$38 25 00.0 &  $-100$& & D \\
G352.7-0.1  & 17 24 20.0   & $-$35 05 00.0 &  $  0$ & & D \\
G354.8-0.8  & 17 32 41.0   & $-$33 40 00.0 &  $-70$ & & D \\
\hline
\hline
\end{tabular}
\end{center}
\tablenotetext{\dagger} {D=1720 MHz line detected by single dish observations.  R=J. Rho's thermal composite class.}
\dummytable\label{D-obs-table}
\end{table}

\begin{table}
\small
\begin{flushleft}
\begin{tabular}{|c|c|c|c|c|c|c|c|}
\multicolumn{8}{c}{\sc Table~\ref{line-fits-table} -- Summary of Spectral line fits for 1720 MHz Masers}
\\
\hline
\hline
Name        & RA(1950)    & Dec(1950)            & $S$ & $\sigma_{rms}$ & $V_{LSR}$ & $\Delta V$  & Notes\\
     & (h m s) & ($\arcdeg$ $\arcmin$ $\arcsec$) & mJy & mJy/beam & km s$^{-1}$ & km s$^{-1}$ & \\
\hline
G32.8-0.1       & 18 49 14.02  & $-$00 14 14.2 & 494  & 4.5 & $-86.1$ & 0.97 & \\
G337.8-0.1      & 16 38 52.15  & $-$46 56 16.1 & 143  &   & $-45$   & ?    & \\
G346.6-0.2      & 17 06 42.23  & $-$40 11 08.6 & 1240 & 11 & $-74.0$ & 1.28  & \\
                &              &               & 200  & & $-76.2$ & 1.74 & 2nd peak \\
                & 17 06 35.53  & $-$40 10 22.2 & 320  & & $-74.3$ & 0.82 & \\
                &              &               & 75.6 & & $-76.2$ & 1.75 & 2nd peak \\
                & 17 06 56.03  & $-$40 09 38.3 & 183  & & $-79.3$ & 1.65 & \\
\hline
\hline
\end{tabular}
\end{flushleft}
\dummytable\label{line-fits-table}
\end{table}

\begin{table}
\small
\begin{flushleft}
\begin{tabular}{|c|c|c|c|c|c|c|}
\multicolumn{7}{c}{\sc Table~\ref{W3-maser-table} -- Summary of 1720 MHz Masers toward W3 and HB3}
\\
\hline
\hline
Name        & RA(1950)    & Dec(1950)            & $S$   & $\sigma_{rms} $ &$V_{LSR}$  & Notes\\
     & (h m s) & ($\arcdeg$ $\arcmin$ $\arcsec$) & mJy   & mJy/beam & km s$^{-1}$ & \\
\hline
W3 Northwest    & 02 21 39.85   & $+$61 54 33.8 & 2215 & 25 & $-38.6$ & \\
W3 H \& D       & 02 21 41.12   & $+$61 53 30.4 & 749  & 25 & $-60.6$ & \\
                & 02 21 41.11   & $+$61 53 30.4 & 550  & 25 & $-63.3$ & 2nd peak \\
                & 02 21 41.42   & $+$61 53 26.0 &  69  &  & $-48.7$ & \\
W3 G            & 02 21 46.89   & $+$61 52 16.0 & 263  & 25 & $-44.7$ & \\
W3OH            & 02 23 16.58   & $+$61 38 58.4 & 1916 & 11 & $-44.7$ & \\
\hline
\hline
\end{tabular}
\end{flushleft}
\dummytable\label{W3-maser-table}
\end{table}

\begin{table}
\small
\begin{flushleft}
\begin{tabular}{|c|c|c|c|c|c|c|c|}
\multicolumn{6}{c}{\sc Table~\ref{detection-table} -- Summary of Maser Detections in this work}
\\
\hline
\hline
Galactic   & Common & SNR   & No.     & $V_{LSR}$   & D (kpc)  & Stokes & B Field  \\
Name       & Name   & Type  & Masers  & km s$^{-1}$ & near, far & V & milliGauss\\
\hline
G32.8-0.1  & Kes 78 & Shell & 1       & $-86.1$     & 5.5, 8.8 & Y & 1.45 \\
G337.8-0.1 & Kes 41 & Shell & 1       & $-45$       & 12.3    & N\tablenotemark{\ddagger} &  \\
G346.6-0.2 &        & Shell & 5       & $-76.0$     & 5.5, 11     & Y & 1.70 \\
\hline
\hline
\end{tabular}
\end{flushleft}
\tablenotetext{\ddagger} {Observations for this remnant were not taken in Stokes V.}
\dummytable\label{detection-table}
\end{table}


\end{document}